\documentclass[a4paper,12pt]{article}
\usepackage{amssymb,amsmath}
\usepackage{a4wide}
\usepackage{epsfig}
\usepackage{graphics}
\usepackage{subfigure}
\usepackage{amsfonts}

\newcommand{\ii}{{\rm i}}
\newcommand{\half}{{\scriptstyle\frac{1}{2}}}
\title{Skyrmion Multi-Walls}
\author{J.\ Silva Lobo\footnote{email address: j.i.silva-lobo@durham.ac.uk}
 \,\, and R.\ S.\ Ward\footnote{email address: richard.ward@durham.ac.uk}
  \bigskip
  \\Deparment of Mathematical Sciences,
  \\Durham University,
  \\Durham DH1 3LE}

\begin{document}
\maketitle
\begin{abstract}
Skyrmion walls are topologically-nontrivial solutions of the Skyrme
system which are periodic in two spatial directions.  We report numerical
investigations which show that solutions representing parallel multi-walls
exist. The most stable configuration is that of the square $N$-wall, which
in the $N\to\infty$ limit becomes the cubically-symmetric Skyrme crystal.
There is also a solution resembling parallel hexagonal walls, but this is
less stable.
\end{abstract}
\vskip 1truein
%\date{\today}
%\noindent {\bf Keywords:} Solitons Monopoles and Instantons; Sigma Models;
%   Global Symmetries. 

\newpage

The Skyrme system, originally introduced as a model of nucleons, is an archetypal
(3+1)-dimensional classical field theory admitting topological soliton solutions.
Much is known about various types of skyrmion solutions, for example: isolated
skyrmions 
in $\mathbb{R}^{3}$, up to relatively high charge \cite{BS02, MS04, BMS07};
a triply-periodic `Skyrme crystal' \cite{K85, KS88, CJJVJ89, MS04};
a doubly-periodic `Skyrme domain wall' \cite{BS98}; and various types of
singly-periodic `Skyrme chains' \cite{HW08}.

The purpose of this note is to investigate static $N$-wall solutions, {\sl ie.}\
the $N>1$ generalization of the single-wall fields discussed in \cite{BS98}.
If one has two (or indeed $N$) well-separated parallel walls, then the force
between them can be made attractive by a suitable relative orientation of the
fields. So one expects there to be solutions representing $N$ walls bound
together, although {\sl a priori} the walls might merge together to form
a single wall.

We investigate this by numerical minimization of the energy, and our main
findings are as follows. There are two obvious shapes for a single wall,
namely square and hexagonal, and it is known \cite{BS98} that the latter has
slightly lower energy than the former. If walls are allowed to attract, then
they do not merge, but remain identifiable as separate parallel walls.
There is a stable bound configuration representing
two parallel hexagonal walls, but this is not the lowest-energy 2-wall state.
For $N\geq2$, the lowest-energy state consists of $N$ parallel square walls (each
one being a square array of half-skyrmions), and as $N\to\infty$ this approaches
the skyrme crystal.

The energy density of a static SU(2)-valued Skyrme field $U(x^j)$ on
$\mathbb{R}^{3}$ is defined to be
\begin{equation}\label{skyrme energy density}
  \mathcal{E}:=-\frac{1}{2}\mbox{tr}(L_i L_i)-
     \frac{1}{16}\mbox{tr}([L_i,L_j][L_i,L_j]),
\end{equation}
where $L_i=U^{-1}\partial U/\partial x^i$, and $x^j=(x,y,z)$ are the spatial
coordinates.
In what follows, let us write $U=\Phi_4+\ii\Phi_j\sigma_j$, where 
$\sigma_j$ are the Pauli matrices, and the real 4-vector field
${\bf\Phi}=(\Phi_1,\Phi_2,\Phi_3,\Phi_4)$ satisfies ${\bf\Phi\cdot\Phi}=1$.

In this note, we deal with configurations which resemble $N$ walls or sheets,
each parallel to the $xy$-plane: so the field is
periodic in $x$ and $y$ (with periods $L_x$ and $L_y$ respectively),
and satisfies the boundary condition
\begin{equation}\label{BC}
 \Phi_4\to\left\{
   \begin{array}{ll}
     1       & \mbox{as $z\to-\infty$}, \\
     (-1)^N  & \mbox{as $z\to\infty$}.
   \end{array}
  \right.
\end{equation}
For $N=1$, and more generally for $N$ odd, one has a domain wall which separates
two vacuum regions, where $\Phi_4=1$ and $\Phi_4=-1$ respectively; for $N$ even,
one has the same vacuum on both sides of the multi-layered sheet.
In the asymptotic region $|z|\gg1$, the three fields $\Phi_j$ are small; and they
satisfy the Laplace equation, since the energy density reduces to
$\mathcal{E}\approx(\partial_i\Phi_j)^2$. Assuming (without loss of generality)
that $L_y\geq L_x$, we see by separating variables that the leading behaviour
as $|z|\to\infty$ is typically $\Phi_j\approx C\sin{(\mu y)}\exp{(-\mu|z|)}$,
where $\mu=2\pi/L_y$. In particular, the fields approach their asymptotic values
exponentially quickly, with a scale determined by the larger of $L_x$ and $L_y$.

The topological charge $Q$ (over a single cell) is
\begin{equation}\label{skyrme charge}
  Q = \int_{T^2\times\mathbb{R}} \mathcal{Q}\, dx\, dy\, dz,
\end{equation}
where
\begin{equation} \label{skyrme charge density}
  \mathcal{Q}= \frac{1}{24\pi^2}\varepsilon_{ijk}\mbox{tr}(L_i L_j L_k)
\end{equation}
is the topological charge density.
We claim that $Q$ is an integer. If $N$ is even, then (\ref{BC}) allows us
to regard ${\bf\Phi}$ as being defined, for topological purposes, on
$T^2\times S^1$; and then $Q$ equals the degree of ${\bf\Phi}$.
If $N$ is odd, then it is not quite so obvious why $Q$ is an integer, but
it follows from the theorem in the Appendix of \cite{HW08}.
The energy $E$ is defined to be
\begin{equation}\label{skyrme energy}
  E:=\frac{1}{12\pi^2}\int_{T^2\times\mathbb{R}}\mathcal{E}\,dx\,dy\,dz,
\end{equation}
and it satisfies the usual Faddeev bound $E\geq Q$.

In what follows, we describe $N$-wall configurations which were found by numerical
minimization of the energy functional $E$.
We used a first-order finite-difference scheme for $E$, with the spatial points
$(x,y,z)$ being represented by a rectangular lattice having lattice spacing $h$,
and we applied conjugate-gradient minimization.
The lattice error in $E$ goes like $h^2$, and we extrapolated the finite-$h$
results for both $E$ and $Q$ to $h=0$. The extrapolated value of $Q$ then gives
a measure of the remaining error, which for the situations described below turns
out to be less than $0.2\%$. The boundary condition (\ref{BC}) was modelled by
imposing $\Phi_4=1$ at $z=-L_z/2$ and $\Phi_4=(-1)^N$ at $z=L_z/2$. As remarked
above, the walls are exponentially
localized in $z$, and so as long as $L_z$ is taken to be large enough, there is
no discernable finite-size effect; a value of $L_z=10+2N$ turns out to be
sufficient for this. In each case, we adjusted the periods $L_x$ and $L_y$
to their optimal size, meaning that the energy-per-cell is made as small as possible.
Numerical minima were randomly perturbed and then re-minimized, as a test of
their stability.  As initial configurations we used the same sort of
`rational map ansatz' as in \cite{BS98}, involving a Weierstrass elliptic function
of $x+\ii y$ (the lemniscatic form to get square symmetry, and the equianharmonic
form to get hexagonal symmetry), together with
a suitable profile function $f(z)$ satisfying $f(-L_z/2)=0$ and $f(L_z/2)=N\pi$.  

The results are consistent with the anticipated general principle that
the lowest-energy configurations are arrays of half-skyrmions.
For an $N$-wall, we expect
that each fundamental cell will contain a multiple of $4N$ half-skyrmions,
and therefore its topological charge $Q$ will be a multiple of $2N$; this
indeed turns out to be the case. As mentioned above, the walls do not merge,
but retain their identity; the location of each wall can be determined by looking
at the locus where $\Phi_4=0$.

The simplest case to describe is the square one, with $L_y=L_x=L$; our results for
$1\leq N\leq5$ are summarized in Table~1, which gives the energy-per-charge
$\widehat{E}$, and the optimal value of $L$, for each $N$.
\begin{table}
\begin{center}
\begin{tabular}{|c|c|c|}
\hline
$N$ & $\widehat{E}$ & $L$ \\
\hline
1 & 1.068 & 4.25 \\
2 & 1.053 & 4.47 \\
3 & 1.048 & 4.54 \\
4 & 1.046 & 4.58 \\
5 & 1.044 & 4.61 \\
\hline
\end{tabular}
\caption{Energy $\widehat{E}$ and cell-size $L$ for the square $N$-wall}\label{Tab1}
\end{center}
\end{table}
Pictures of the $N=1$ and $N=2$ cases are presented in Figure~1, together with a
plot of the energy data in Table~1.
\begin{figure}[htb]
\begin{center}
\includegraphics[scale=1.0]{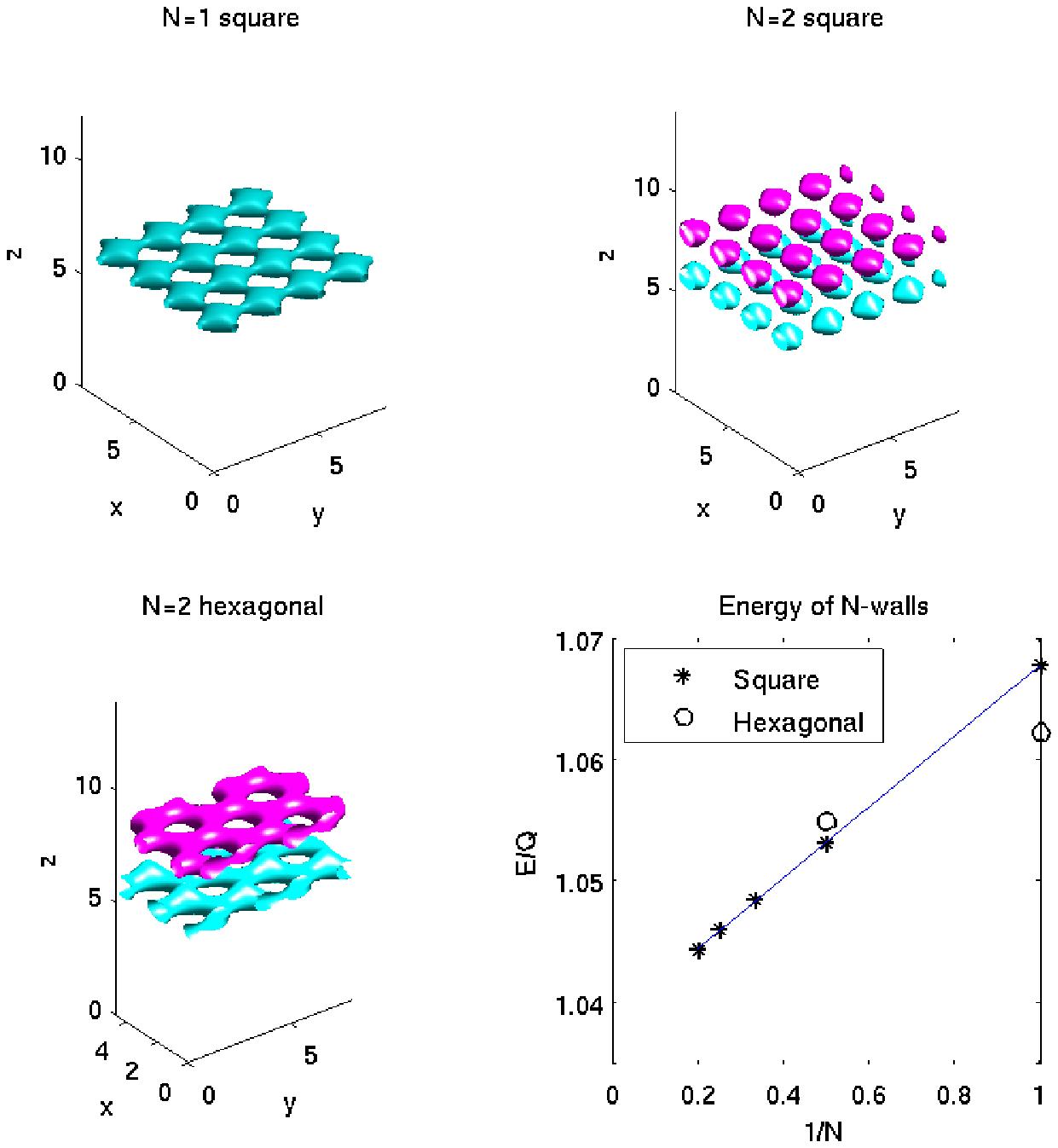}
\caption{Energy densities of the square 1-wall, square 2-wall, and hexagonal
  2-wall; and plot of the energy $\widehat{E}$ for the square $N$-wall
($1\leq N\leq5$) and hexagonal $N$-wall ($1\leq N\leq2$). \label{Fig1}}
\end{center}
\end{figure}
Let us first comment on the data. The normalized energy $\widehat{E}$
of the square $N$-wall is surprisingly close to having a $1/N$-dependence
(although there is no obvious reason why this should be so); and
extrapolating on this basis gives $\widehat{E}\approx1.039$
in the $N\to\infty$ limit. This is very close to the energy of the (triply-periodic)
skyrme crystal, a cubic array in which each fundamental cube contains eight
half-skyrmions: its energy-per-charge, computed using the method described above,
is $\widehat{E}=1.038$.
Further support for the claim that the square $N$-wall tends to the skyrme crystal
as $N\to\infty$ comes from looking at the symmetries of the field ${\bf\Phi}$.
These include, for example, the translations
\begin{eqnarray*}
x\mapsto x+\half L_x &\Rightarrow&
   (\Phi_1,\Phi_2,\Phi_3,\Phi_4)\mapsto(-\Phi_1,-\Phi_2,\Phi_3,\Phi_4),  \\
y\mapsto y+\half L_y &\Rightarrow&
   (\Phi_1,\Phi_2,\Phi_3,\Phi_4)\mapsto(\Phi_1,-\Phi_2,-\Phi_3,\Phi_4),  \\
z_{p}\mapsto z_{p+1} &\Rightarrow&
   (\Phi_1,\Phi_2,\Phi_3,\Phi_4)\mapsto(\Phi_1,-\Phi_2,\Phi_3,-\Phi_4),
\end{eqnarray*}
where the third translation (in $z$) denotes moving from the $p^{{\rm th}}$
wall to the $(p+1)^{{\rm st}}$ wall. These are exactly the same as the
translation symmetries of the
skyrme crystal \cite{MS04}. The values for the optimal cell-length $L=L_x=L_y$
are consistent with their approaching $L=4.7$ as $N\to\infty$, this being the
cell-size of the skyrme crystal (and similarly the distance between each parallel
pair of walls is approximately $4.7/2$, as one would expect).

Each 3-dimensional plot in Figure~1 is an isosurface of the energy density
$\mathcal{E}$, namely where $\mathcal{E}$ equals 0.6 times its maximum value.
For the square case, the plots are over four fundamental cells. One clearly
sees square arrays of half-skyrmions. Observe that, for $N=2$, the half-skyrmions
are aligned in the $z$-direction; the same is true for $N>2$.

Let us turn now to the case of hexagonal symmetry. For ease of computation,
we follow the same scheme as in \cite{BS98}, namely taking $L_y=\sqrt{3}L_x$
and fitting two fundamental parallelograms into the corresponding rectangle.
Each such rectangle, of each wall, contains eight half-skyrmions, as
is seen in the hexagonal 2-wall picture of Figure~1. For $N=1$, the energy
of the hexagonal arrangement is $\widehat{E}\approx1.062$, less than that
of the corresponding square case \cite{BS98}; but for $N\geq2$, the hexagonal
arrangement is less efficient than the square one, and (depending on the values
of $L_x$ and $L_y$) it is either a local minimum of the energy functional, or
it is unstable. There is a local-minimum hexagonal 2-wall solution with energy
$\widehat{E}\approx1.055$, which is only very slightly
(less than $0.2\%$) higher than that of the square 2-wall. Its energy density is
depicted in Figure~1; one feature to note is that the two walls are not aligned
in the $z$-direction, but are offset. If $L_x$ and $L_y$ are allowed to change
so that the relation $L_y=\sqrt{3}L_x$ no longer holds, then this solution
becomes unstable, and changes into the square 2-wall.

An isolated skyrmion of charge $Q\geq3$ typically has a polyhedral-shell structure,
analogous to carbon fullerenes, and it may be viewed as constructed from a section
of the hexagonal 1-wall (graphene), with the insertion of defects to create a spherical
shell \cite{BS98, MS04}. There has also been an investigation \cite{MP01} of
the possibility of constructing skyrmions as multi-walled spherical shells, with
the `shell material' consisting of a double or triple wall. For the cases
that were examined in \cite{MP01}, either the walls coalesced, or one
obtained a structure which resembled a shell-like part of the skyrme crystal.
The findings reported above are consistent with this; in particular,
multiple hexagonal walls appear to be rather unstable, and therefore unsuitable
for constructing shells. But it does raise
the possibility of stable high-charge skyrmions constructed as shells of
square multi-wall material, or equivalently as hollow chunks of skyrme
crystal, and this would be worth investigating further.

\bigskip\noindent{\bf Acknowledgment.}
Support from the UK Science and Technology Facilities Council, through the
Rolling Grant `Particles, Fields and Spacetime', is gratefully acknowledged.


\begin{thebibliography}{99}

\bibitem{BS02}
R~Battye and P~M~Sutcliffe, Skyrmions, fullerenes and rational maps.
           {\it Rev Math Phys} {\bf14} (2002) 29--85.

\bibitem{MS04}
N~S~Manton and P~M Sutcliffe, {\it Topological Solitons.}
    (Cambridge University Press, 2004)

\bibitem{BMS07}
R~Battye, N~S~Manton and P~M~Sutcliffe, Skyrmions and the $\alpha$-particle
      model of nuclei. {\it Proc~Roy~Soc~Lond~A} {\bf463} (2007) 261--279.

\bibitem{K85}
I~Klebanov, Nuclear matter in the Skyrme model.
        {\it Nucl Phys B} {\bf262} (1985) 133--143.

\bibitem{KS88}
M~Kugler and S~Shtrikman, A new Skyrmion crystal.
    {\it Phys Lett B} {\bf208} (1988) 491--494.


\bibitem{CJJVJ89}
L~Castillejo, P~S~J~Jones, A~D~Jackson, J~J~M~Verbaarschot and A~Jackson,
        Dense Skyrmion systems {\it Nucl Phys A} {\bf501} (1989) 801--812.

\bibitem{BS98}
R~Battye and P~M~Sutcliffe, A Skyrme lattice with hexagonal symmetry.
   {\it Phys Lett B} {\bf416} (1998) 385--391.

\bibitem{HW08}
D~Harland and R~S~Ward, Chains of skyrmions.
        {\it JHEP} {\bf12} (2008) 093.

\bibitem{MP01}
N~S~Manton and B~M~A~G~Piette, Understanding skyrmions using rational maps.
        {\it Prog Math} {\bf201} (2001) 469--480.

\end{thebibliography}
\end{document}